\documentclass[12pt]{article}

\pdfoutput=1

\usepackage{cite}

\usepackage{amssymb,latexsym}

\usepackage{epstopdf}

\usepackage{graphics,epsfig}

\usepackage{color}

\let\a=\alpha \let\b=\beta \let\g=\gamma \let\d=\delta \let\e=\epsilon
  \let\th=\theta  \let\k=\kappa
\let\l=\lambda \let\m=\mu \let\n=\nu \let\x=\xi \let\p=\pi 
\let\s=\sigma   \let\f=\phi  
\let\vp=\varphi 
        \let\Th=\Theta 
\let\X=\Xi  \let\S=\Sigma  \let\Y=\Psi
 
\let\la=\label  
  
\def\nn{\nonumber} \def\bd{\begin{document}} \def\ed{\end{document}}
\def\ds{\documentstyle} \let\fr=\frac \let\bl=\bigl \let\br=\bigr
\let\Br=\Bigr \let\Bl=\Bigl
\let\bm=\bibitem
\let\na=\nabla
\def\tU{{\widetilde U}}
\let\pa=\partial \let\ov=\overline
\def\ie{{\it i.e.\ }}
\newcommand{\be}{\begin{equation}}
\newcommand{\ee}{\end{equation}}
\def\ba{\begin{array}}
\def\ea{\end{array}}
\def\ft#1#2{{\textstyle{{\scriptstyle #1}\over {\scriptstyle #2}}}}
\def\fft#1#2{{#1 \over #2}}
\def\F#1#2{{ F_{#1}^{(#2)} }}
\def\cF#1#2{{ {\cal F}_{#1}^{(#2)} }}

\def\R{{\bf R}}
\def\sst#1{{\scriptscriptstyle #1}}
\def\oneone{\rlap 1\mkern4mu{\rm l}}
\def\e7{E_{7(+7)}}
\def\td{\tilde}
\def\wtd{\widetilde}
\def\im{{\rm i}}
\def\bog{Bogomol'nyi\ }
\newcommand{\ho}[1]{$\, ^{#1}$}
\newcommand{\hoch}[1]{$\, ^{#1}$}
\newcommand{\bea}{\begin{eqnarray}}
\newcommand{\eea}{\end{eqnarray}}
\newcommand{\ra}{\rightarrow}
\newcommand{\lra}{\longrightarrow}
\newcommand{\Lra}{\Leftrightarrow}
\newcommand{\ap}{\alpha^\prime}
\newcommand{\bp}{\tilde \beta^\prime}
\newcommand{\cB}{{\cal B}}
\newcommand{\cO}{{\cal O}}
\newcommand{\vecx}{\vec{x}}
\newcommand{\vecy}{\vec{y}}
\newcommand{\vecp}{\vec{p}}
\newcommand{\vecq}{\vec{q}}
\newcommand{\tr}{{\rm tr} }
\newcommand{\Tr}{{\rm Tr} }
\newcommand{\NP}{Nucl. Phys. }

\newcommand{\cL}{{\cal L}}
\newcommand{\cA}{{\cal A}}
\newcommand{\cT}{{\cal T}}
\newcommand{\cD}{{\cal D}}
\newcommand{\cH}{{\cal H}}

\def\sst#1{{\scriptscriptstyle #1}}
\def\0{{\sst{(0)}}}
\def\1{{\sst{(1)}}}
\def\2{{\sst{(2)}}}
\def\3{{\sst{(3)}}}
\def\4{{\sst{(4)}}}
\def\5{{\sst{(5)}}}
\def\6{{\sst{(6)}}}
\def\7{{\sst{(7)}}}
\def\8{{\sst{(8)}}}
\def\9{{\sst{(9)}}}
\def\p{{\sst{(p)}}}
\def\q{{\sst{(q)}}}
\def\ve{\varepsilon}
\def\vf{\varphi}
\def\F{\Phi}
\def\wg{\wedge}

\def\thb{\bar{\theta}}
\def\Thb{\bar{\Theta}}
\def\barp{\bar{p}}
\def\barq{\bar{q}}
\def\barc{\bar{c}}
\def\bard{\bar{d}}
\def\e{\epsilon}

\def \bi{\bibitem}
\def \la {\label}

\def \l {\lambda}
\def\foot{\footnote}
\def \tl  {{\tilde \l}}
\def \sql {{\sqrt \l}}
\def \adss {$AdS_5 \times S^5$\ }
\newcommand{\rf}[1]{(\ref{#1})}
\def \ov {\over}

\def\th{\theta}
\def\Th{\Theta}
\def\vth{\vartheta}
\def\btheta{{\bar\theta}}
\def\ttheta{{{\tilde\theta}}}
\def\bttheta{{{\bar\ttheta}}}
\def\vth{\vartheta}

\def\ra{\rightarrow}
\def\N{\nabla}
\def\F{{\cal F}}
\def\uM{\underline{M}}
\def\uA{\underline{A}}
\def\uN{\underline{N}}
\def\uP{\underline{P}}
\def\ua{\underline{a}}
\def\ub{\underline{b}}
\def\uc{\underline{c}}
\def\ud{\underline{d}}
\def\ue{\underline{e}}
\def\uf{\underline{f}}
\def\ui{\underline{i}}
\def\uj{\underline{j}}
\def\uk{\underline{k}}
\def\ul{\underline{l}}
\def\ual{\underline{\alpha}}
\def\ube{\underline{\beta}}
\def\um{\underline{m}}
\def\un{\underline{n}}
\def\up{\underline{p}}
\def\uq{\underline{q}}
\def\ur{\underline{r}}
\def\us{\underline{s}}
\def\umu{\underline{\mu}}
\def\unu{\underline{\nu}}
\def\ula{\underline{\l}}
\def\uka{\underline{\k}}
\def\usi{\underline{\s}}
\def\urh{\underline{\r}}
\def\cc{\circ}
\def\eqv{\equiv}

\def\ni{\noindent}

\def\Ep{E^{{}^{(+)}}}
\def\Em{E^{{}^{(-)}}}

\def\Mp{M^{{}^{(+)}}}
\def\Mm{M^{{}^{(-)}}}

\def \ha{{1\ov 2}}

\def\r{\rho}

\def\Y{{\rm Y}}
\def\X{{\rm X}}
\def\tY{\tilde{\rm Y}}
\def\tX{\tilde{\rm X}}
\def\dY{\dot{\rm Y}}
\def\dX{\dot{\rm X}}

\def \J {\mathcal{J}}
\def \del {\partial}

\def\dF{\dot{F}}
\def\dG{\dot{G}}
\def\df{\dot{f}}
\def \E {{\cal E}}
\def \S {{\cal S}}
\def \J {{\cal J}}

\def\ms{\mathcal{S}}
\def\mj{\mathcal{J}}
\def\soj{\fr{\ms}{\mj}}
\def \R {{\bf R}}
\def \om {\omega}
\def \bE {\bar E}
\def \x {{\cal X}}

\def \bi{\bibitem}
\def \la {\label}

\def \l {\lambda}
\def\foot{\footnote}
\def \tl  {{\tilde \l}}
\def \sql {{\sqrt \l}}
\def \adss {$AdS_5 \times S^5$\ }
\def \ov {\over}

\def \varpi {{\rm w}}

\def\thb{\bar{\theta}}
\def\Thb{\bar{\Theta}}
\def\mb{\bar{\m}}
\def\ab{\bar{\a}}
\def\zb{\bar{z}}
\def\psib{\bar{\psi}}
\def\barp{\bar{p}}
\def\barq{\bar{q}}
\def\barc{\bar{c}}
\def\bard{\bar{d}}
\def\e{\epsilon}
\def\wb{\bar{w}}
\def\lb{\bar{\l}}
\def\Jb{\bar{J}}
\def\Nb{\bar{N}}
\def\Zb{\bar{Z}}
\def\pab{\bar{\pa}}

\def\At{\tilde{A}}
\def\Bt{\tilde{B}}
\def\Ct{\tilde{C}}
\def\Dt{\tilde{D}}
\def\Et{\tilde{E}}
\def\Ft{\tilde{F}}
\def\Gt{\tilde{G}}
\def\Ht{\tilde{H}}
\def\Mt{\tilde{M}}
\def\Rt{\tilde{R}}
\def\at{\tilde{a}}
\def\bt{\tilde{b}}
\def\ct{\tilde{c}}
\def\dt{\tilde{d}}
\def\et{\tilde{e}}
\def\ft{\tilde{f}}
\def\htil{\tilde{h}}
\def\gt{\tilde{g}}
\def\mt{\tilde{\mu}}
\def\nt{\tilde{\nu}}
\def\pht{\tilde{\f}}

\def\rht{\tilde{\rho}}

\def\asth{\hat{*}}
\def\phh{\hat{\phi}}

\def\bA{{\bf A}}

\def\ola{\overleftarrow}
\def\ora{\overrightarrow}
\def\alt{\tilde{\a}}

\def\eh{\hat{e}}
\def\eph{\hat{\e}}
\def\ph{\hat{p}}
\def\alh{\hat{\a}}
\def\beh{\hat{\b}}
\def\gah{\hat{\g}}
\def\Fh{\hat{F}}
\def\muh{\hat{\m}}
\def\nuh{\hat{\n}}
\def\thh{\hat{\th}}
\def\rhh{\hat{\r}}
\def\xih{\hat{\xi}}
\def\phh{\hat{\phi}}
\def\chh{\hat{\chi}}
\def\dh{\hat{d}}
\def\ih{\hat{i}}
\def\jh{\hat{j}}
\def\kh{\hat{k}}
\def\hh{\hat{h}}
\def\nh{\hat{n}}
\def\Nh{\hat{N}}
\def\deh{\hat{\d}}
\def\wh{\hat{w}}
\def\lah{\hat{\l}}
\def\Ah{\hat{A}}
\def\Ch{\hat{C}}
\def\Omh{\hat{\Omega}}

\def\xh{\hat{x}}
\def\Nah{\hat{\N}}
\def\Rh{\hat{R}}

\def\ps{\rlap{\, /}\;\,p }
\def\ks{\rlap{\, /}\;\,k }

\def\gym{g_{YM}}

\def\adot{\dot{a}}
\def\bdot{\dot{b}}
\def\bpa{\bar{\pa}}

\def\pr{\prime}
\def\ssk{\medskip}
\def\clb{\color{blue}}
\def\clr{\color{red}}
\def\clv{\color{violet}}
\def\clg{\color{green}}

\begin{document}

\overfullrule=0pt
\parskip=2pt
\parindent=12pt
\headheight=0in \headsep=0in \topmargin=0in
\oddsidemargin=0in

\vspace{ -3cm}
\thispagestyle{empty}

 \vspace{0.1cm}

\setcounter{equation}{0}
\setcounter{footnote}{0}
\setcounter{section}{0}

\begin{center}

{\Large\bf  Dimensional reduction to hypersurface of foliation}

\vskip 0.8cm

 \vspace{.5cm}


\vspace{0.5cm}
I. Y. Park
\\
[7mm]
{\it Department of Physics, Hanyang University \\
Seoul 133-791, Korea}\\

\vspace{0.3cm}

{\it Center for Quantum Spacetime, Sogang University\\
Shinsu-dong 1, Mapo-gu, 121-742 South Korea \\
}
\vspace{0.1cm}
and

\vspace{0.1cm}
{\it Department of Applied Mathematics,
Philander Smith College 
                               \\
Little Rock, AR 72223, USA \\
inyongpark05@gmail.com
}

\end{center}

 \vspace{0.1cm}

 \begin{abstract}

When the bulk spacetime has a foliation structure, the collective dynamics of the hypersurfaces should
reveal certain aspects of the bulk physics. 
The procedure of reducing the bulk to a hypersurface, called ADM reduction, was implemented in \cite{Park:2013iqa} where the 4D Einstein-Hilbert 
action was reduced along the radial reduction. In this work, reduction along the angular 
directions is considered {with a main goal to firmly establish the method of dimensional reduction to a hypersurface of foliation.} We obtain a theory on a 2D plane (the $(t,r)$-plane) and 
observe that novel and elaborate boundary effects are crucial for the consistency of the reduction. 
The reduction leads to a 2D interacting quantum field theory. We comment on its 
application to black hole information physics.

\end{abstract}
\newpage

\section{Introduction}

When the spacetime has a foliation structure and a high degree of symmetry, certain aspects of 
the bulk physics may admit a holographic "dual" description. The holographic description
should be realized through the dynamics of the hypersurfaces: the leaves of the foliation. 
For this realization, a technique - which we called ADM reduction - for reducing the bulk to a selected hypersurface is needed. Using the ingredients in \cite{Sato:2002kv}, the reduction was proposed in \cite{Hatefi:2012bp} and further developed in \cite{Park:2013iqa}
 where the 4D Einstein-Hilbert action was reduced along the radial direction. (A subsequent development can be found, e.g., in \cite{Park:2013bma}.)
Building on these previous works {and paying a careful attention to the boundary terms}, we carry out angular ADM reduction of Schwarzschild black hole
spacetime here {with a main goal of establishing the method of dimensional reduction to a hypersurface of foliation.} 

This work was motivated by an explicit realization of the "holographic duality" and its application
to black hole physics. Our prime goal of relatively long-term is to develop a framework
optimal for analyzing scattering around a black hole as well as related issues.

Below, we reduce the 4D Einstein-Hilbert action to a 2D action in the $(t,r)$-plane by carrying out
the ADM reduction along the angular directions of the Schwarzschild black hole. There are several reasons
for reducing the action to 2D through the ADM procedure. (The first two reasons are good for for standard reduction as well.) The first reason is obviously quantization-related: it is not known how to second-quantize the 4D or 3D gravity but the second quantization is essential for the multi-particle scattering. One can bypass the 3D or 4D quantization issue by reducing the theory to 2D where the metric becomes topological and non-fluctuating. As we will see below, the 4D Einstein-Hilbert action reduces to an interacting theory of scalars in which the 2D metric serves as a background.

The second reason for 2D being preferred is a matter of simplicity. Even if the 4D Einstein-Hilbert action were quantizable, it would still be advantageous to reduce it to 2D since that would allow a simpler, although narrower and more limited, description of the dynamics. Using a reduced theory means that only a particular sector of the moduli space of the original 4D theory is described, {and one should be cautious in projecting the results to the higher dimensions.}\footnote{On the other hand, if the 4D theory has a genuine pathology such as the information problem, it must manifest in a reduced theory as well; it simply cannot be true that the information problem present in 4D suddenly disappears in a reduced theory.} This is a relatively small price compared with what one gains: a tool optimized in many ways for analyzing black hole information at the quantum field theory level.

\begin{figure}
\centerline{
\begin{minipage}[b]{8cm}
             \epsfxsize=8cm
              \epsfbox{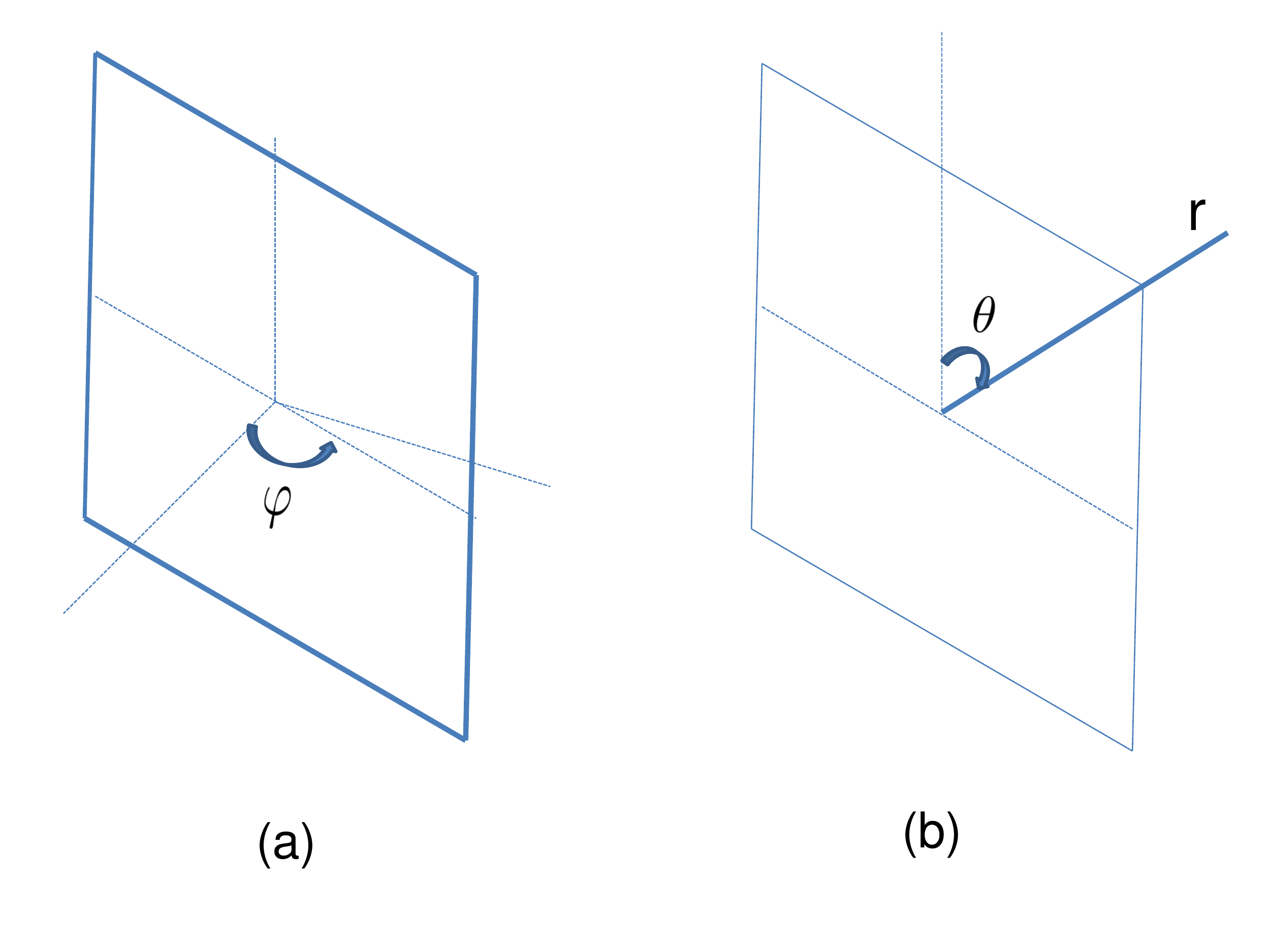}
      \end{minipage}
      }
\caption{reduction to $(t,r,\th)$-plane (a) followed by reduction to $(t,r)$-plane (b)}
\label{fig}
\end{figure}

Thirdly, following the ADM reduction procedure has one great advantage over the conventional dimensional reduction when it comes to describing the physics of a hypersurface in the curved background such as the Schwarzschild background.
The ADM procedure offers a very clear physical interpretation as 
 shown in Fig.1. The figure (a) shows reduction to the space of the fixed value of $\vp$, one of the angular coordinates, $(t,r,\th,\vf)$. The reduced theory should describe the dynamics of the $(t,r,\th)$-plane.
With the $\th$-direction subsequently reduced, one gets a 2D theory in the $(t,r)$-plane as shown in Fig.1 (b).

At the technical level, only the reduction where the shift vector is set to zero will be considered.
The ADM reduction has another distinguishing feature in addition to offering a clear physical interpretation. It turns out that
the selected hypersurface acts like a virtual boundary of the bulk.
When there is no boundary, total derivative terms do not matter. However, they do 
matter in general in a case in which the bulk is reduced to a hypersurface. This surface
effectively becomes a boundary of the bulk.
Because of this, one should be cautious when dealing with the surface terms in the action, and we show that
there are elaborate boundary terms that lead to consistent reduction: {the role of the virtual boundary terms is crucial for consistent reduction when one tries to reduce along a non-isometry direction.}

\begin{figure}
\centerline{
\begin{minipage}[b]{8cm}
             \epsfxsize=8cm
              \epsfbox{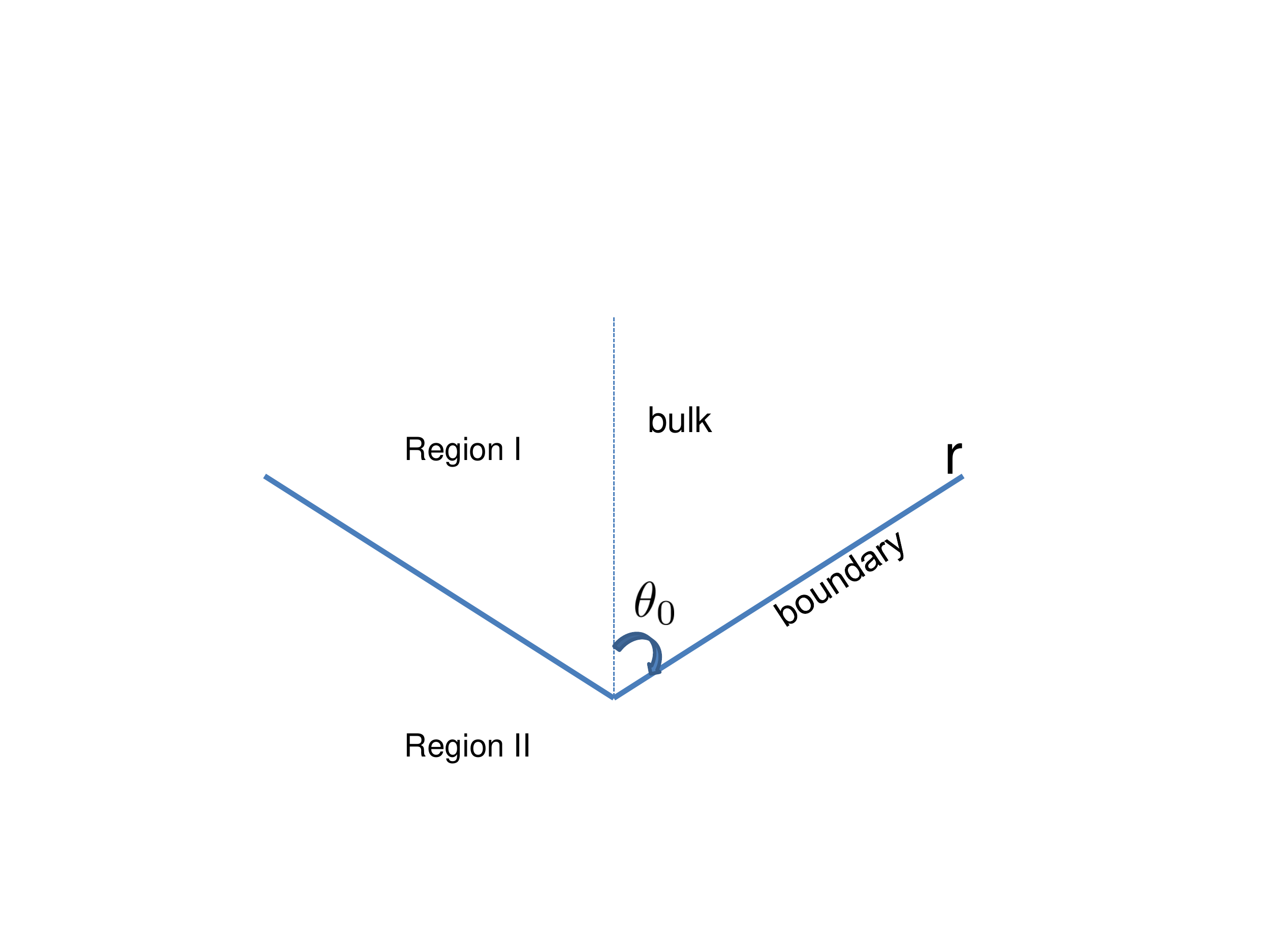}
      \end{minipage}
      }
\caption{virtual boundary}
\label{fig}
\end{figure}

\vspace{.3in}
The remainder is organized as follows. In the next section we start with the 4D Einstein-Hilbert action, and reduce one of the angular coordinates, $\vf$. It is the reduction that is depicted in Fig.1 (a). Because we narrow down to the sector where the shift vector is absent and $\vf$ is an isometry direction, the procedure is essentially
that of the stand $S^1$ reduction.\footnote{The standard reduction, e.g., on $S^2$ can be found in \cite{Lechtenfeld:1992rt}.
} The subsequent reduction of $\th$ direction has several
subtleties, among which is the novel emergence of the virtual boundary.
Once the boundary is present, one must choose the boundary terms appropriately.\footnote{
In the context of an event horizon \cite{Carlip:2002be}, the relevance of boundary effects were discussed, e.g., in \cite{Park:2001zn} and \cite{Padmanabhan:2009vy}. (We thank Mu-In Park for pointing out \cite{Padmanabhan:2009vy}. Our result here is more radical in the sense that any hypersurface can potentially induce boundary effects in the context of ADM reduction.)
}
With the boundary effects properly taken into account, 
we show in section 3 that the resulting theory is a two-dimensional interacting
 theory of two scalars. {One novelty is that the 2D action has coordinate dependent coefficients. Some of them arise from the virtual boundary terms; the others come from the fact that the action is reduced in the Schwarzschild background as opposed to, say, a flat background.\footnote{{Suppose one expands the 4D action around the Schwarzschild background and then reduces. The reduced action will have coordinate-dependent coefficients. The coordinate dependence that we will see in this work has the same origin.}}}  A related novelty is the appearance of a "cosmological function" term. The scalars are {\em intrinsic} degrees of freedom of the original
4D Einstein-Hilbert action. 
We conclude with future directions; in particular we comment on applications to the black hole physics.

\section{3D Field equations with boundary terms}

 Reduction of $\vf$ direction is straightforward for several reasons: the $\vf$ is an isometry direction and we consider only the case in which the shift vector is absent. (As a matter of fact, it can be viewed as a gauge-fixing.) 
 As we will see, a coupled system of the metric and a scalar field will arise after the reduction.
Further reduction to 2D is more involved but can be achieved by carefully handling
the boundary contributions.
The resulting 2D theory is an interacting system of two scalars in a fixed metric background - ``the reduced 2D Schwarzschild".

We start with 4D Einstein-Hilbert action,
\bea
S=\int d^4 x\sqrt{-g}\, R^\4  \la{EH}
\eea
The 4D coordinate $x^{\umu}$ consists of the azimuthal angle $\vf$ and the 3D coordinate $x^\m$:
\bea
x^{\umu}=(x^\m,\vf),\quad x^\m=(x^i,\th)\;\;i=0,1\;\;\mbox{where}\;\;x^i\equiv (t,r) 
\eea
Reduction to 3D without the shift vector is straightforward; let us take the ansatz
\be
ds^2_4=h_{\m\n}(x) dx^\m dx^\n+e^{2 \rhh(x)}d\vf^2 
\la{ansq}
\ee
The reduction can be carried out by closely following the technique of the standard dimensional reduction on a circle.  
Nevertheless, the reduction should be viewed as ADM reduction
instead, and this is  
the only nontrivial aspect of the reduction.

For $\vf$ reduction, this viewpoint is a matter of convenience to a large extent.
However, as for the subsequent reduction along $\th$, which is not an isometry direction, the ADM viewpoint is dictated by the consistency of reduction: if one starts with the 3D action in the context of the standard $S^1$ reduction, the reduction along $\th$ would lead to inconsistency. 
The inconsistency originates precisely from the boundary effects, and there is no room
for the boundary effects in the standard $S^1$ reduction. Once the effects of the virtual boundary are taken into account, the same ansatz becomes consistent in the ADM context as we will show.
In other words, reduction along a non-isometry direction comes to have a meaning in the ADM reduction 
setup where the hypersurface acts as a virtual boundary in the manner unraveled below.

\vspace{.2in}

By taking the following ansatz 
\be
ds^2_4=h_{\m\n}(x) dx^\m dx^\n+e^{2 \rhh(x)}d\vf^2 
\la{ans}
\ee
and reducing along $\vf$, the reduced 3D action takes
\be
\int d^3 x\cL^\3=\int d^3 x e^{\rhh} \sqrt{-h}\left[R^\3-2\Nah^2 \rhh-2(\pa \rhh)^2 \right]
\la{redLag}
\ee
The second and third terms combined are a total derivative, and the action can be
rewritten as
\bea
\int d^3x\cL^\3 
                &=& \int d^3x \sqrt{-h}\Big(e^{\rhh} R^\3-2 \Nah^2 e^{\rhh}\Big)
\la{redLag2}
\eea
Let us gauge fix the metric according to
\bea
h_{\m\n}=\left(
\begin{array}{ccc} 
 h_{tt} & h_{tr} & 0  \\
 h_{rt} & h_{rr} & 0\\
  0 & 0 & h_{\th\th}
\end{array}
\right)
\eea
Further reduction to 2D will be carried out below by taking the ansatze
\bea
ds^2_3 &=& \g_{ij}(t,r)dx^i dx^j+e^{2\xi(t,r)}d\th^2  \nn\\
  \rhh &=& \r_0(\th)+\r(t,r)
\la{ans2q}
\eea
As emphasized, the total derivative term in \rf{redLag2} will play an 
important role in the reduction to 2D. {The background $\r_0(\th)$ deserves remarks. The {\em fields} in the reduced 2D action will only depend on $(t,r)$. However, the action will have $\th$-dependent coefficients, a trace of the 4D Schwarzschild background. }

\vspace{.2in}

A total derivative term can be omitted in the absence of a boundary. In the presence of a boundary, a total
derivative yields surface contributions.
Things become subtle if one sets out to find equations of motions of the {\em reduced} theory. The surface terms can now be important because they are
likely to change the field equations of the reduced theory. Therefore, in the presence of a genuine
boundary, the importance of the surface terms is evident. The question
is whether a hypersurface in ADM reduction acts like a genuine boundary.
Our analysis indicates that the answer is affirmative: the hypersurface acts virtually 
as a boundary and should be treated as such.

Let us focus on the total derivative term\footnote{The naive action that results from \rf{redLag2} by dropping the total derivative term $\Nah^2 e^{\rhh}$,
\bea
S_{3D}  &=& \int d^3x  \sqrt{-h}\;e^{\rhh} R^\3  \la{nact}
\eea
leads to a problem as we will point out below.
},
\bea
\int d^2x\int_0^{\th_0} d\th \sqrt{-h}\;\Nah^2 e^{\rhh}
\la{3Dextraf}
\eea
with the bulk space taken as the region I in Fig.2. $\th_0=\pi$ corresponds to the original entire bulk spacetime. {As we will see now, the form \rf{3Dextraf} will give a hint for the precise forms of the boundary terms that are determined in  \rf{redLag3} below.} The relevance of splitting the space into regions I,II is not obvious a priori; it just seems to be the peculiar way in which ADM reduction generates a virtual boundary.  
Applying Gauss theorem, one gets (see, e.g., \cite{tool})
\bea
&& \int d^3 x\sqrt{-h}\;\Nah^2 e^{\rhh} 
 =\int  d^2x \sqrt{-h_{\th\th}h^\2}\;\pa^\th e^{\rhh}|_{\th_0}  \la{surface}
\eea
where $|_{\th_0}$ indicates that the entire integrand has been evaluated at the fixed angle $\th$. The 2 by 2 matrix $h_{ij}^\2$ is  
\bea
h^\2_{ij}\equiv \left(
\begin{array}{cc}
 h_{tt} & h_{tr}  \\
 h_{rt} & h_{rr} \\
\end{array}
\right)
\eea
Although the integrand on the right-hand side of \rf{surface} is not a total $\pa_\th$-derivative (since $h^{\th\th}, h^\2$ depend on $\th$ as well before reduction), it can be rewritten as a bulk term. However, there are some subtleties: the bulk expression is not unique.
It turns out that the bulk expressions that leads to consistent reduction is the following\footnote{
The bulk expression is not unique because eq.\rf{surface} can be expressed another way.
Treating all the factors in the integrand democratically, it can be written, e.g., as
\bea 
  \int  d^2x \int_0^{\th_0}d\th{ \sqrt{-h^{\th\th}h^\2}}\;\pa_\th^2 e^{\rhh}
\eea
instead of the left-hand side of \rf{redLag5}. (In other words, $h_{\th\th}$ is not evaluated at $\th=\th_0$ either.)
Or a more precise expression is
\bea 
  \int  d^2x \int_0^{\th_0}d\th { \sqrt{-h^{\th\th}h^\2}}\;e^{-\xi}\pa_\th^2 e^{\rhh}+\cdots
\eea
Here $(\cdots)$ represents two kinds of the terms. (Such terms are also present in \rf{redLag5}, of course.) There are terms that vanish once the action is reduced to 2D. Some terms will be of the Gibbons-Hawking type. The other kind
is the term that comes from evaluating the integrand at $\th=0$; it does not contribute to the 3D field equations. {The presence of the types of terms in $(\cdots)$ will be illustrated in the appendix and they will be omitted in the remaining discussion of the main body.}
}
:
\bea 
\int  d^3x{ \sqrt{-h^\2}}\;\sqrt{(h_{\th\th}|_{\th_0})}\;h^{\th\th}\pa_\th^2 e^{\rhh}  
  \Rightarrow \int  d^3x{ \sqrt{-h^\2}}\;e^{-\xi}\pa_\th^2 e^{\rhh}
  \la{redLag5}
\eea
where the right arrow indicates the fact that the expression on the left-hand side can be
replaced by the one on the right-hand side without affecting the final result, say, of the 2D action in the next section.

 Eq.\rf{redLag5} gives only a partial guide to the precise forms of the boundary terms. It is eventually the consistency of the reduction that determines the precise forms of the boundary terms.  
It turns out that the action that leads to consistent reduction is give by
\bea
S_{3D}  
      &=& \int d^3x \sqrt{-h}\;e^{\rhh} R^\3
{ +2\int  d^3x\sqrt{-h^\2}\; e^{-\xi}\pa_\th^2 e^{\rhh}
-2\int  d^3x\sqrt{-h^\2}\; e^{-\xi_0}\pa_\th^2 e^{\rhh_0}
} \nn\\
&& \hspace{2.5in}+\mbox{surface terms} 
\la{redLag3}   
\eea
where $\xi_0$ and $\rhh_0$ ($\equiv \r_0$) are fixed functions defined in \rf{xirho2dsol} below.
"Surface terms" include a Gibbons-Hawking type \cite{Gibbons:1976ue} term, and will be explicitly addressed below.
Note that the 3D covariance is broken by the boundary effect {(and it will be broken further by the Schwarzschild background below)}.
The presence of the third term in \rf{redLag3} is more curious; it is the consistency of the reduction that dictates its presence as we will see shortly.
The origin of the precise forms of the boundary terms in \rf{redLag3} is not entirely clear other than 
that it should be due to the boundary effects.  
In spite of the fact that obtaining \rf{redLag3} involves subtle and elaborate boundary effects,
the Lagrangian \rf{redLag3} will enjoy the hallmark property of Kaluza-Klein reduction: a solution
of \rf{redLag3} can be embedded as the corresponding solution of the original action, and the same is true for the 2D action obtained below.

Let us turn to the field equations. The  $h^{ij}$-variation leads to\footnote{
The naive action \rf{nact} leads, instead, to
\be
e^{\rhh}(R^\3_{ij}-\fr12 R^\3 h_{ij})-\Nah_i \Nah_j e^{\rhh}+h_{ij}\Nah^2 e^{\rhh}
=0
\la{heompartijn}
\ee
which, in turn, is reduced to
\bea
G_{ij}-\N_i\N_j\xi-\N_i\xi\N_j\xi-\N_i\N_j\rhh-\N_i\rhh\N_j\rhh 
  \la{2D3eomredn}  
\eea
\[
+\g_{ij}\Big(\N^2\xi+(\N\xi)^2 +\N^2\rhh+(\N\rhh)^2+\N\xi \cdot \N\rhh\Big)
+{ \g_{ij}e^{-2\xi-\rhh}\pa_\th^2 e^{\rhh}}
=0 
\]
instead of \rf{2D3eomred} below. The difference between the two Einstein equations lies only in the last terms.
As far as we can see, it is impossible to construct a consistently reduced action with the last term
as given in \rf{2D3eomredn}. The boundary terms in \rf{redLag3} were obtained through an elaborate analysis in order
to put the last term in \rf{2D3eomredn} into the form given in \rf{2D3eomred} and to keep the ``reduced Schwarzschild configuration", \rf{g2dsol} and \rf{xirho2dsol}, as a solution at the same time.
}
\be
e^{\rhh}(R^\3_{ij}-\fr12 R^\3 h_{ij})-\Nah_i \Nah_j e^{\rhh}+h_{ij}\Nah^2 e^{\rhh}
{ -h_{ij}e^{-2\xi}\pa_\th^2 e^{\rhh}+h_{ij}e^{-\xi-\xi_0}\pa_\th^2 e^{\rhh_0}}=0
\la{heompartij}
\ee
where $h_{\th\th}$ has been replaced by $e^{2\xi}$ in anticipation of the reduction 
in the next section.
 As for the $(\th\th)$-component, one gets
\bea
 e^{\rhh}(R^\3_{\th\th}-\fr12 R^\3 h_{\th\th})
-\Nah_\th \Nah_\th e^{\rhh}+h_{\th\th}\Nah^2 e^{\rhh}
 =0
\la{heomth}
\eea
Next let us consider the $\rhh$ field equation. The $\rhh$-variation of the boundary term in \rf{redLag5} 
, once reduced to 2D, can be removed by the corresponding (Gibbons-Hawking type) boundary terms in \rf{redLag3}.
To see this note that the $\rhh$-variation of the second term in \rf{redLag3} yields
\bea
2\int  d^3x\sqrt{-h^\2}\; e^{-\xi}\pa_\th [ e^{\rhh}\pa_\th \d \rhh+\d\rhh \pa_\th e^{\rhh} ]
\la{illus}
\eea
This terms do not contribute to the 2D equations of motion and therefore will not be considered further; more details of the analysis can be found in the appendix.
With this, one gets, for the $\rhh$ field equation,
\bea
e^{\rhh} R^\3=0  \la{rhhfe}
\eea

\section{Reduction of 3D field eqs to 2D}

Above we have obtained the field equations for the metric and $\rhh$; they are
\rf{heompartij}, \rf{heomth} and \rf{rhhfe}.\footnote{
Due to the various extra terms mentioned below \rf{heomth}, carrying out the reduction
at the level of variation would be a more streamlined approach. For simplicity, however, we only 
consider the bulk field equations here.    
}
The goal of this section is first to reduce those 3D field equations to the $(t,r)$-plane, and then to obtain the 2D Lagrangian that produces the reduced field equations. We take the following ansatz
\bea
ds^2_3 &=& \g_{ij}(t,r)dx^i dx^j+e^{2\xi(t,r)}d\th^2  \nn\\
  \rhh &=& \r_0(\th)+\r(t,r)
\la{ans2}
\eea
The $(ij)$-component of the 3D Einstein equation \rf{heompartij} becomes
\bea
G_{ij}-\N_i\N_j\xi-\N_i\xi\N_j\xi-\N_i\N_j\rhh-\N_i\rhh\N_j\rhh 
  \la{2D3eomred}  
\eea
\[
+\g_{ij}\Big(\N^2\xi+(\N\xi)^2 +\N^2\rhh+(\N\rhh)^2+\N\xi \cdot \N\rhh\Big)
+{ \g_{ij}e^{-\xi-{ \rhh}-\xi_0}\pa_\th^2 e^{\rhh_0}}
=0 
\]
where $\r_0,\xi_0$ are fixed functions: $e^{\xi_0}\equiv r, e^{\r_0}\equiv r\sin\th$.
The $h^{\th\th}$ field equation \rf{heomth} reduces to
\bea
-\fr12 R^\2+\N^2\rhh+(\N \rhh)^2=0 \la{hththred}
\eea
The $\rhh$ field equation \rf{rhhfe} reduces to
\be
 R^\2-2\N^2\xi-2(\N \xi)^2 =0
\la{2rhheom}
\ee
The field equations \rf{2D3eomred}, \rf{hththred} and \rf{2rhheom} are expected to
admit (and indeed they do admit) the following "reduced Schwarzschild solution",\footnote{Since $\th$ is no longer a coordinate for the 2D system but a parameter instead, the fact that the solution for $\rhh$ depends on $\th$ is perfectly ok.}
\be
\g_{0ij}={\mathrm{diag}}(-f(r),1/f(r)),\qquad f(r)=1-\a/r
\la{g2dsol}
\ee
\be
e^{\xi}=e^{\xi_0}=r,\qquad e^{\rhh}=e^{\r_0}=r\sin\th
\la{xirho2dsol}
\ee
This is by design because the original system has been reduced to the degrees of freedom that fluctuate
around the configuration \rf{g2dsol} and \rf{xirho2dsol}.

It is relatively straightforward to construct an action:
\bea
\cL^\2 &=& \sqrt{-\g}e^{\xi+\tilde{\r}}\left(R^\2+2\pa_i \xi \pa^i \tilde{\r} \right)
-{ 2}{ \sqrt{-\g}\;e^{-\xi_0}\pa_\th^2 e^{\r_0}}  \nn\\
\la{2Dlageq}
\eea
where again $\r_0,\xi_0$ are defined as in \rf{xirho2dsol}.
The second term is of boundary origin, and it is a "cosmological function" term. {(It may potentially have some interesting implications.)}
Let us check the field equations that follow from \rf{2Dlageq}.
The 2D metric equation is
\bea
e^{\xi+\tilde{\r}}\left(G^\2_{ij}-\g_{ij}(\pa \tilde{\r} \cdot \pa\xi)+2 \pa_i \tilde{\r} \pa_j\xi \right)-\N_i \N_j e^{\xi+\tilde{\r}}+\g_{ij}\N^2 e^{\xi+\tilde{\r}}
{+\g_{ij}e^{-\xi_0}\pa^2_\th e^{\r_0}}=0  \nn\\
\la{Eincc}
\eea
This can be simplified to yield:
\bea
G_{ij}-\N_i\N_j\xi-\N_i\xi\N_j\xi-\N_i\N_j{\tilde \r}-\N_i{\tilde \r}\N_j{\tilde \r} 
\eea
\[
+\g_{ij}\Big(\N^2\xi+(\N\xi)^2 +\N^2\tilde{\r}+(\N\tilde{\r})^2+\N\xi \cdot \N\tilde{\r}\Big)
+{ \g_{ij}e^{-\xi-{\tilde{\r}}-\xi_0}\pa_\th^2 e^{\rhh_0}}
=0 
\]
This is identical to \rf{2D3eomred} upon setting $\tilde{\r}= \rhh$.
Note that the second term in \rf{2Dlageq} which is of boundary origin has contributed. 
Once it is
present, it with the third term contributes to the $\xi$ field equation. 
The $\xi$ field equation is
\bea
&& R^\2-2\N^2{\tilde \r}-2(\pa {\tilde \r})^2=0
\la{xieomccr}
\eea
The $\r$ field equation is
\be
e^\xi \left(R^\2-2\N^2\xi-2(\N \xi)^2 \right)=0
\la{2D1eomq}
\ee
Therefore the action \rf{2Dlageq} reproduces all the three field equations, \rf{2D3eomred}, \rf{hththred} and \rf{2rhheom}. 

Since the 2D metric is non-dynamical, it can be set to the reduced Schwarzschild solution.
{The Lagrangian will have coordinate-dependent coefficients and a systematic analysis would be required to determine issues such as renormalizability.} (The issue of renormalizability will be pursued elsewhere.)
The action should then be supplemented with the metric field equation, which 
serves as a constraint.

\section{Conclusion}

In this work, we have carried out ADM reduction of 4D Einstein-Hilbert action along the two angular directions of the Schwarzschild solution, and constructed the 2D action \rf{2Dlageq} that admits the reduced Schwarzschild solution \rf{g2dsol} and \rf{xirho2dsol}. The boundary effects were crucial for the consistency of the reduction. A ``cosmological function" term appeared in the 2D action; whether it has a deeper meaning remains to be seen.
We believe that this framework of the ADM reduction should have a potential for 
providing a large class of "dual" pairs of gravity/non-gravity theories. A similar method applied to
IIB supergravity led to gravity/gauge pairs \cite{Sato:2002kv}\cite{Hatefi:2012bp}.
A mechanism for obtaining a non-abelian SYM was sketched in \cite{Hatefi:2012bp}.

\begin{figure}
\centerline{
\begin{minipage}[b]{8cm}
             \epsfxsize=8cm
              \epsfbox{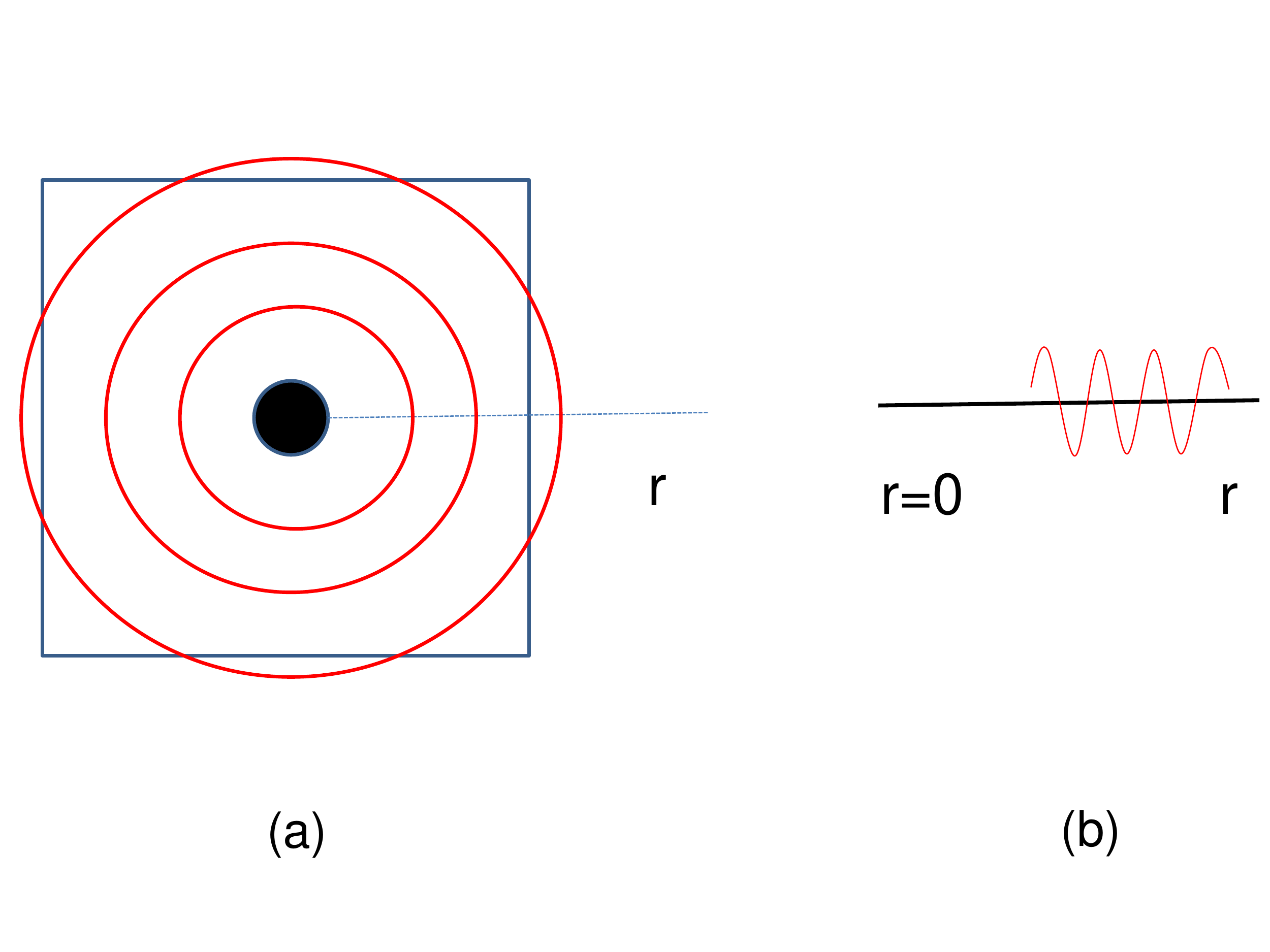}
      \end{minipage}
      }
\caption{scattering around a black hole in (a) 3D and (b) 2D setups}
\label{fig}
\end{figure}

As stated in the introduction, our primary motivation for this work was to develop a quantum field theory
framework to tackle the black hole information in the context of the 4D Einstein-Hilbert action.
It has been noted in \cite{Park:2013rm} that the potential presence of a mechanism for information bleaching 
may provide a solution to the information problem. (Also, the potential presence of "blackening" process should be important as pointed out in the same reference. See \cite{Hayward:2005gi} for a related discussion.) 
Consider a particle incoming towards a black hole. The particle will generate jets, if it has a QFT interaction, as it approaches the black hole. (The jet production will be analogous to the Bremsstrahlung process \cite{Heckler:1997jv}\cite{Page:2007yr} or jet quenching. ) By measuring the jets, it should be possible to recover much of the information about the incoming particle. One of the key questions is whether the jet production process can be related to an information-bleaching mechanism of a certain kind. If so, it implies that
some of the information does not enter the event horizon when the particles enter.\footnote{
It is not difficult to come up up with a situation that could potentially be viewed as certain
information-bleaching. As a particle approaches a black hole, it would produce jets, and the amount
of jets will depend on the initial velocity. The information on the initial velocity might be bleached into the surrounding space of the BH, especially into the horizon and its vicinity.
Analyzing such a process would require a well-designed quantum field theory framework, and we believe that the ADM reduction
developed here should be one such framework.
}

One of the future directions is to analyze scattering around the Schwarzschild black hole
by using the 2D setup obtained in the main body (see Fig. 3). (See \cite{Almheiri:2013wka} for a recent quantitative discussion.) We will report on this elsewhere.

\vspace{.5in} \ni {\bf Acknowledgments}

\ni I thank A. Nurmagambetov for valuable discussions throughout the project.

\newpage

\appendix

\renewcommand{\theequation}{A.\arabic{equation}}
 \setcounter{equation}{0}

\section{Some details on the boundary terms}

Here we present a more detailed analysis of the boundary terms by taking the example of \rf{illus}:
\bea
2\int  d^3x\sqrt{-h^\2}\; e^{-\xi}\pa_\th [ e^{\rhh}\pa_\th \d \rhh+\d\r \pa_\th e^{\rhh} ]
\la{illusq}
\eea
This can be re-written
\bea
&& 2\int  d^3x\; \pa_\th\Big[ \sqrt{-h^\2}\; e^{-\xi} ( e^{\rhh}\pa_\th \d \rhh+\d\rhh \pa_\th e^{\rhh} )\Big] \nn\\
&&-2\int  d^3x \Big[\pa_\th (\sqrt{-h^\2}\; e^{-\xi})\Big] ( e^{\rhh}\pa_\th \d \rhh+\d\rhh \pa_\th e^{\rhh} )
\la{illusq}
\eea
The first term is a Gibbons-Hawking boundary term: the term with $\pa_\th \d \rhh$ vanishes after the reduction since $\d\rhh=\d \r(t,r)$ and the term with $\d \rhh$ enforces the Dirichlet boundary condition on $\rhh$. The second term contributes to the 3D field equations. It, however, vanishes upon the reduction 2D since 
\bea
\Big[\pa_\th (\sqrt{-h^\2}\; e^{-\xi})\Big]=0
\eea 
as dictated by the $\th$ independence of $h^\2$ and $\xi$ of the reduction ansatz.

\end{document}